\documentclass[twocolumn,secnumarabic,amssymb,nobibnotesaps,prd]{revtex4-1}
\usepackage{amsmath}
\usepackage{graphicx}

\makeindex

\begin{document}

\title{Probing high-energy interactions of atmospheric and astrophysical neutrinos}

\author{Spencer R. Klein}

\address{Nuclear Science Division \\
Lawrence Berkeley National Labaoratory\\
Berkeley, CA, 94720 USA, and \\
Physics Department, \\
University of California, Berkeley \\
Berkeley, CA, 94720 USA \\
srklein@lbl.gov}

\begin{abstract}

Astrophysical and atmospheric neutrinos are important probes of the powerful accelerators that produce cosmic-rays with EeV energies.   Understanding these accelerators is a key goal of neutrino observatories, along with searches for neutrinos from supernovae, from dark matter annihilation, and other astrophysics topics.   Here, we discuss how neutrino observatories like IceCube and future facilities like KM3NeT and IceCube-Gen2 can study the properties of high-energy (above 1 TeV) neutrino interactions.  This is far higher than is accessible at man-made accelerators, where the highest energy neutrino beam reached only 500 GeV. In contrast, neutrino observatories have observed events with energies above 5 PeV - 10,000 times higher in energy - and future large observatories may probe neutrinos with energies up to $10^{20}$ eV.    These data have  implications for both Standard Model measurements, such as of low Bjorken$-x$ parton distributions and gluon shadowing, and also for searches for beyond Standard Model physics.  This chapter will review the existing techniques and results, and discuss future prospects. 

\end{abstract}
\maketitle

\section{Introduction}
\label{sec:sources}

Compared to neutrinos produced at accelerators, natural (atmospheric and astrophysical) neutrinos have both advantages and disadvantages.   Astrophysical neutrinos have been observed with energies above 5~PeV \cite{Aartsen:2016xlq}, more than 10,000 times that available at accelerators.  They are studied at much longer baselines (production to detection distance) than  at accelerators.  The Earth's diameter is 12,800 km, while the longest existing accelerator baseline is 810 km, from Fermilab to the Nova detector in northern Minnesota.    The larger baselines enable qualitatively new types of measurements, such as the observation of neutrino absorption in the Earth.   

The observed neutrino flux in neutrino telescopes includes contributions from conventional atmospheric neutrinos, prompt atmospheric neutrinos and an astrophysical flux, as discussed elsewhere \cite{Halzen:2008zz}. Conventional atmospheric neutrinos come  from the decay of $\pi^\pm$, $K^\pm$ and $K^0$ produced in cosmic-ray air showers, while prompt atmospheric neutrinos arise from the decays of similarly produced charmed hadrons.  Conventional neutrinos are mostly $\nu_\mu$ (and $\overline\nu_\mu$), while the prompt decays produce 50\% $\nu_\mu$ and 50\% $\nu_e$.  

Finally, the still-poorly understood astrophysical flux is generally parameterized as a single power law, $d\Phi_\nu/dE_\nu = \Phi_0 (E_\nu/100\  {\rm TeV}) ^{-\gamma}$ where $\gamma$ is the spectral index, 100 TeV is the so-called pivot point, and $\Phi_0$ is the flux normalization, with an equal flux for each flavor.   Different analyses have found different values for $\gamma$, but it should be  between 2.0 and 3.0.  

The $\nu:\overline\nu$ ratio is assumed to be $1:1$,  but the actual ratio is still unmeasured.  This ratio can depend on the mechanism by which neutrinos are produced in the source.  Proton-proton interactions lead to a different ratio than proton-photon interactions.   Unfortunately, we have very little ability to measure this ratio.  At energies above about 5 PeV, the Glashow resonance (from $\overline\nu_e e \rightarrow W^-$)  \cite{Glashow:1960zz} can help select $\overline\nu_e$, and at energies below about 10 TeV, differences in the charged-current (CC) reaction neutrino and antineutrino inelasticity (fraction of the neutrino energy transferred to the target nucleus) can provide some statistical separation between neutrinos and antineutrinos.  However,  we have only observed one or two neutrinos with energies above 5 PeV, and these two techniques are currently only of limited application.   In the rest of this chapter, we will lump neutrinos and antineutrinos together, unless explicitly noted.  

As our data improve, it is likely that the assumed single power law and fixed $\nu:\overline\nu$ ratio may become inadequate.  As we will see, in most experimental analyses, it is not possible to separate uncertainties in the beam from uncertainties in the interaction properties, so the flux parameterization plays a role in cross-section and inelasticity measurements.

At the energies considered here, above about 1 TeV \cite{Klein:2018waq}, standard-model neutrino oscillations within the Earth are negligible, so atmospheric neutrinos reach the detector with their initial  flavor ratios.   Astrophysical neutrinos travel much larger distances before being observed, so should have oscillated enough to reach an effective equilibrium, with the $\nu_e:\nu_\mu:\nu_\tau$ ratio close to $1:1:1$.  This ratio is quite insensitive to the flavor ratio at the production point, with models based on pion decay (initial flavor ratio $1:2:0$), neutron decay (initial flavor ratio $1:0:0$), muon energy loss (initial flavor ratio $0:1:0$)  \cite{Kashti:2005qa} or with muon acceleration (initial flavor ratio $1:1:0$)  \cite{Klein:2012ug} predicting similar results \cite{Aartsen:2015ivb}.  Some exotic models with beyond-standard-model interactions or neutrino sources could alter these ratios.  An anomalous ratio would be visible in the inelasticity study discussed below, but we will not further consider this possibility here.

Natural neutrinos do have limitations.   The beam intensity is limited, so a large detector (order 1 km$^3$) is required to collect a large sample of neutrinos with TeV or higher energies.  So, detector granularity is necessarily limited.  This limits their event reconstruction capabilities, particularly for neutrino flavor identification.  The neutrino beam spans a broad energy range, and its energy spectrum, flavor composition and $\nu:\overline\nu$ ratio are not well known.  Uncertainties on these beam characteristics lead to systematic errors in any analysis.   

Ideally, analyses would  separately identify $\nu_e$, $\nu_\mu$ and $\nu_\tau$.  Unfortunately, this is not generally possible.   Neutrino telescopes usually classify events into one of two topologies: tracks, and cascades (electromagnetic or hadronic showers) \cite{Halzen:2010yj}.  Tracks are either backgrounds from downward-going cosmic-ray interactions, or from $\nu_\mu$ or $\overline\nu_\mu$ CC  interactions, generally outside the detector.  One subclass, starting tracks, come from neutrino interactions within the detector.  In these events,  a track emerges from a cascade in the detector.

Cascades are produced by CC interactions of $\nu_e$ and $\overline\nu_e$ and neutral-current (NC) interactions of all flavors.    The differences between electromagnetic and hadronic cascades are small, and we cannot currently separate these two classes of events.  One proposed observable may offer some statistical separation: late (delayed) light in the shower, from neutrons capture on proton and from muon decays \cite{Li:2016kra}.  This late light should be present only in hadronic showers.  If the backgrounds can be adequately understood, this could be a useful tool for measuring the $\nu_e$ flux.

Charged-current $\nu_\tau$  interactions can fit into either category, depending on how the produced $\tau$ decays.  In the most striking, but as-yet unobserved topology, a $\tau$ with an energy of a few PeV will travel a few hundred meters and decay, producing a `double-bang' - one cascade when the neutrino interacts, and a second when the $\tau$ decays \cite{Learned:1994wg}.  At lower energies, the two bangs merge and $\nu_\tau$ interactions look like tracks or cascades.   Most  $\tau$ decays lead to cascades, but $\tau\rightarrow\mu\overline\nu_\mu\nu_\tau$ (and the antiparticle equivalent) generate track-like events.  In all $\tau$ decays,  the energy carried off by the outgoing $\nu_\tau$ and, if present, other neutrinos, reduces the energy that is visible in the detector.  IceCube has made several studies of flavor identification, generally by comparing the rates of tracks and cascades, or by searching for tau-like signatures.  These analyses have limited statistical power, so current data only lightly constrains some of the exotic models for astrophysical neutrino flavor ratios, such as all $\nu_e$ or all $\nu_\mu$ \cite{Aartsen:2015ivb,Aartsen:2015knd}.   

Neutrino interactions occur via $W^\pm$ or $Z^0$ exchange, so  probe the distribution of quarks in nuclei at low Bjorken$-x$ values.    Alternately, neutrino studies can look for deviations from the standard model, {\it i. e.} new physics which generates new types of neutrino interactions.  New interactions have two effects.  No matter what final state is produced, they should increase the cross-section.  New interactions are likely to lead to final states that look different from the standard model products.  Changes in final state can be observed by studying contained neutrino interactions.  The simplest observable is the CC inelasticity, a measure of how a $\nu_\mu$ partitions its energy between the target nucleus and the outgoing muon.   This requires separately determining the energy of the hadronic cascade and of the produced muon.  The contained event and absorption strategies are complementary: absorption is sensitive to all new interactions, while searches for new final states can have a higher sensitivity, but are less inclusive. 

A cross-section increase leads to increased absorption in the Earth.   Charged current $\nu_e$ and $\nu_\mu$ lead to complete absorption, since the produced leptons lose their energy and the neutrinos essentially disappear.   $\nu_\tau$ are more complex.  Because they are heavy, $\tau^\pm$ lose energy relatively slowly, so high-energy $\tau^\pm$ may decay before losing all of their energy.  These decays produce $\nu_\tau$ (and possibly $\nu_\mu$ and $\nu_e$), albeit with less energy than the initial $\nu_\tau$  \cite{Beacom:2001xn,Dutta:2005yt}.  This is known as tau regeneration.  Because atmospheric $\nu_\tau$ are so rare, this effect is unimportant for most current measurements, but it may be significant as absorption studies  progress toward higher energies where astrophysical neutrinos dominate. Neutral-current interactions are similar to $\nu_\tau$ CC interactions, in that the neutrinos only lose a fraction of their energy.  Because of this, absorption must be treated as a two-dimensional problem (energy out vs. energy in), rather than just as a simple function of neutrino energy.  The overall spectrum must therefore be considered.

\section{Theory}
\label{sec:theory}

Weak interactions are discussed in many textbooks \cite{Collins1989}, so we present only a very brief review.   At energies above ~50 GeV, neutrinos primarily interact via vector boson exchange.  CC interactions with a nucleus $A$, like $\nu_l + A\rightarrow l^+ + X$ occur via $W^\pm$ exchange, while NC interactions occur via $Z^0$ exchange, $\nu_x + A\rightarrow \nu_x + X$.   In both cases, the incident neutrino transfers a fraction of its energy (an average of 50\% at low energies,  decreasing to an average of 20\% at very high energies) to a struck quark, with the outgoing lepton carrying away the remaining energy.   The coupling constants are small, so lowest-order calculations are relatively accurate.  For neutrino energies $k\ll M_Z,M_W$ the cross-section rises linearly with photon energy.  At higher energy, above about 1 TeV, the increase moderates, and the cross-section goes roughly as $E_{\nu}^{0.3}$. 

\begin{figure}
\centerline{\includegraphics[width=0.5\textwidth]{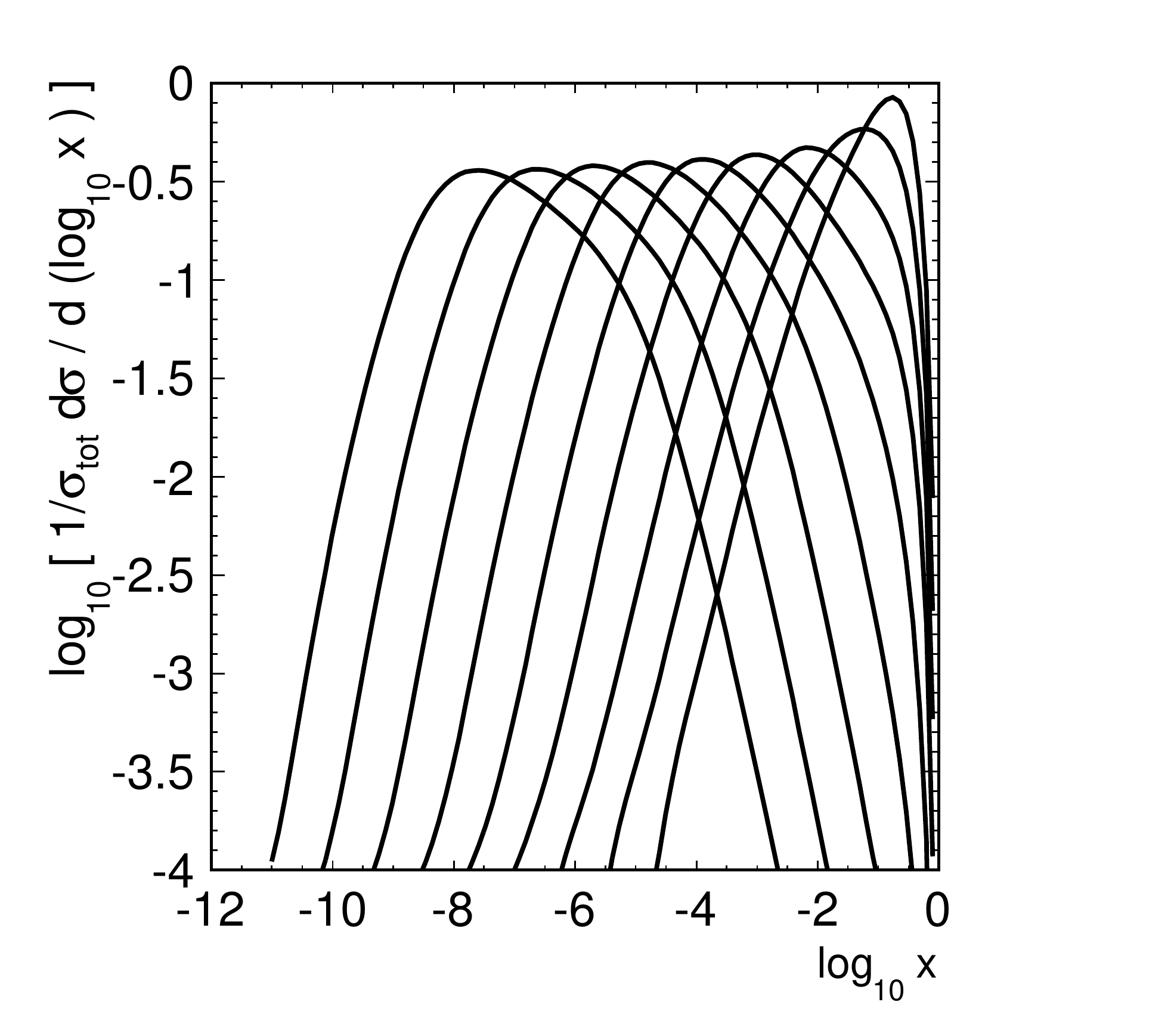}}
\caption{The relative contributions of quarks with different Bjorken$-x$ values (on the x-axis) to neutrino-nucleon interactions for energies with (from the right) $10^4$ GeV to $10^{12}$ GeV, moving to the left in factor-of-10 steps. From Ref. \cite{Connolly:2011vc}.}
\label{fig:Bjorken}
\end{figure}

The range of Bjorken$-x$ values probed by neutrino interactions depends on the neutrino energy.  Figure \ref{fig:Bjorken} shows the relative contributions of quarks with different Bjorken$-x$ values  to neutrino interactions at different energies.  Optical Cherenkov detectors probe a fairly narrow range of Bjorken$-x$, $x>{\rm few} 10^{-3}$, but future experiments using radio-detection can reach neutrino energies up to $10^{20}$ eV, so reach $x$ values down to $10^{-6}$.   The momentum transfers, $Q^2 \approx M_{W,Z}^2$, are high, so these interactions will probe a previously inaccessible kinematic region.   At these low $x$ values, the parton densities are very high, and the neutrino cross-section and mean inelasticity are sensitive to possible non-linear parton dynamics in this region
\cite{CastroPena:2000fx}.

Although lowest order (LO) calculations describe the data reasonably well,  newer calculations go beyond leading order.  The so-called CSMS  next-to-leading order (NLO) calculations uses the DGLAP formalism for parton evolution, using matched (to NLO) parton distributions \cite{CooperSarkar:2011pa}.   These calculations use DGLAP evolution to evolve the parton distributions into unmeasured regions.  They account for heavy quarks  by using general-mass-variable flavor number schemes.  The authors have carefully evaluated the uncertainties, and estimate an uncertainty of less than 5\% up to $E_\nu =1$ PeV.  At higher energies, the cross-section probes quarks with increasingly small Bjorken$-x$ values, and the uncertainty increases significantly.    At very high energies, unitarity constraints might limit the parton distributions, reducing the cross-sections \cite{Goncalves:2015fua}.

There are a few other issues to consider.   They are not very significant with our current level of precision, but they are likely 5-10\% effects that will become important with increasingly precise measurements.  One is isospin; for $W^\pm$ exchange, neutrinos interact differently with protons and neutrons because of their different quark content.  Most of the nuclear targets in the Earth have similar numbers of protons and neutrons, and current analyses treat the Earth as an isoscalar target.   But, this is not completely accurate. 

Another  is nuclear shadowing; nuclear ($A>1$) parton distributions are not merely the sum of their constituent protons and neutrons.   At low Bjorken$-x$, nuclear shadowing reduces the effective number of quarks and gluons.  The MinerVA collaboration has observed nuclear shadowing in nuclear targets with 5-50 GeV neutrinos for Bjorken$-x< 0.1$.   They observe about a 15\% effect for lead and 5\% for iron, but the shadowing should increase with increasing neutrino energy (and hence decreasing Bjorken$-x$) \cite{Mousseau:2016snl}.

Thirdly, neutrinos may also interact electromagnetically.  A high energy neutrino may undergo a short-lived quantum fluctuation into a lepton plus a $W$; the lepton may then interact with the Coulomb field of a target nucleus \cite{Seckel:1997kk,Alikhanov:2015kla,Gauld:2019pgt}, leading to a reaction $\overline\nu_e e\rightarrow W^-\gamma$, where the $W$ decays to $d\overline{u}$, $s\overline{c}$, or $l^-\overline\nu$.  Because the Coulomb field scales as $Z^2$, the cross-section cannot be treated as being per-nucleon.  Any analysis needs to account for the nuclear composition of the Earth.  These interactions begin to be significant at energies of about 50 TeV.  At an energy of 1 PeV, the cross-section for these interactions is about 8\% (25\%) of the charged-current cross-section in oxygen (iron), so can affect absorption for neutrinos passing through the Earth's core.   At higher energies, the cross-section drops off slowly.    

Finally, the previously mentioned `Glashow resonance \cite{Glashow:1960zz}  produces a very large peak in the $\overline\nu_e$ cross-section at an energy of 6.4 PeV (where the $\nu$-e center of mass energy is $M_W$).  There, the cross-section rises by more than two orders of magnitude, and even short chords through the Earth become strongly absorptive.   Because of the natural width of the W boson, this process dominates over charged-current and neutral-current interactions for energies between about 5 PeV and 10 PeV.   At these energies, the neutrino flux is almost entirely astrophysical, so the unknown astrophysical $\nu_e:\overline\nu_e$ ratio has a strong effect on any cross-section measurement in this energy range.   

In addition to altering the neutrino cross-sections, all of these interactions can affect the character of the interactions, changing the inelasticity distribution.  

Beyond-standard-model physics can also impact the neutrino cross-section.  Especially at higher energies, cross-section measurements can be sensitive to new physics.   Here, we consider a few representative possibilities:  leptoquarks, additional dimensions, sphalerons, and supersymmetry.  

Leptoquarks are particles that couple to both leptons and quarks; it is easy to see how they can increase the neutrino cross-section.  When the neutrino-quark center of mass energy reaches the leptoquark mass, the resulting resonance increases the cross-section  by orders of magnitude \cite{Romero:2009vu}.   In different models, the leptoquark may couple to different quarks and leptons, with reactions allowed either within the same generation ({\it i. e.} with $u/d$ quarks, electrons and $\nu_e$ only), or with mixing allowed between generations.  Many models have already been probed by CERN's Large Hadron Collider, but there are regions of phase space where neutrino telescopes have higher sensitivity \cite{Becirevic:2018uab}.  Future radio-detection experiments will be capable of pushing mass exclusion regions much higher. 

If there are additional spatial dimensions in our universe, they must be rolled up at very short distances, and essentially invisible to us.  However, when the momentum transfer between the neutrino and nuclear targets gets large enough (roughly momentum transfer $p >\hbar/R_d$, where $R_d$ is the length scale for the dimension), then the additional dimensions become accessible, and the cross-section rises dramatically, altering the angular distributions of upward-going muons from neutrinos \cite{AlvarezMuniz:2002ga,AlvarezMuniz:2001mk}. 
 
Sphalerons  are created by non-perturbative topological effects on $SU(2)$ groups, as is found in QCD.   They mediate changes in the Cherns-Simons winding number \cite{Cohen:1993nk}.  They could explain phenomena like the creation of a baryon asymmetry in the early universe.  A sufficiently energetic neutrino may be able to produce a reaction which changes the Cherns-Simons winding number.   Different calculations have found different values of the required effective mass $E_{\rm sph}$, but they may be in the 10 TeV range.  When the neutrino-nucleon center-of-mass energy is larger than $E_{\rm sph}$, the cross-section increases drastically.  $E_{\rm sph}$ around 10 TeV corresponds to neutrino energies above $5\times 10^{17}$ eV.   One can also look for sphalerons by studying the character of observed neutrino interactions.  Ellis, Sakurai and Spannowsky found that, for $E_{\rm sph} > 9$ TeV, IceCube provides tighter constraints than ATLAS at the Large Hadron Collider \cite{Ellis:2016dgb}.
 
Although supersymmetry does not lead to large changes in the neutrino cross-section, it can produce some new (and as-yet unseen) topologies in models where the next-to-lightest supersymmetric particle (NLSP) is charged and  long-lived.  NLSPs are produced in pairs, and, because of their lifetime and high mass, can travel hundreds of miles through the Earth \cite{Albuquerque:2006am}.  The two tracks will gradually separate as they propagate, due to their initial relative transverse momentum and due to multiple scattering \cite{Albuquerque:2009vk}.  In neutrino telescopes, they may be observed as a pair of upward-going tracks.   Similar events may also be produced in some Kaluza-Klein theories that include additional rolled-up spatial dimensions.  The competing Standard Model background comes from a single air shower which produces two neutrinos that interact in a detector.  The rate for this is small, but it may be visible in next-generation neutrino detectors \cite{vanderDrift:2013zga}.   A search in IceCube found no evidence of a signal \cite{Kopper:2015rrp}.   

\section{Neutrino Absorption/disappearance Measurements}

\subsection{Early developments}

The idea of neutrino absorption in the Earth was first discussed by Volkova and Zatsepin in 1974, using neutrinos from an accelerator \cite{Volkova:1974xa}.  They proposed to use neutrinos to measure the Earth's density profile, {\it i. e.} performing neutrino tomography of the Earth.  Other authors expanded on this concept, sometimes with a view to identifying oil or gas deposits, or dense metal ores \cite{DeRujula:1983ya}.  Wilson also considered tomography, introducing the idea of using astrophysical neutrinos as a beam \cite{Wilson:1983an}, and other authors followed this path \cite{Reynoso:2004dt,Ralston:1999fz}.   The first specific proposal came from Crawford et al. in 1995, using DUMAND as a detector \cite{Hank}.   As neutrino telescopes became a reality,  more realistic studies were performed, based on measurements of the atmospheric neutrino flux \cite{GonzalezGarcia:2007gg,Kotoyo}.  

Ohlsson and Winter proposed a similar study, but using matter-enhanced neutrino oscillations as a probe \cite{Ohlsson:2001fy}, as a means to probe search for petroleum deposits.  They concluded that this was a very difficult endeavor, since this involves much lower energy neutrinos and a different means of neutrino disappearance.

In 2002, Hooper proposed using neutrino absorption as a means to measure the neutrino cross-section, rather than the Earth's properties  \cite{Hooper:2002yq}.  He proposed using a simple metric: the ratio of upgoing to downgoing neutrinos.  This approach has an appealing symmetry but, because it does not account for the differing path lengths through the Earth of upgoing neutrinos, or asymmetries in atmospheric properties, it does not make the most efficient use of the data.  Binning neutrinos in terms of zenith angle and energy is a more efficient use of the available data \cite{Borriello:2007cs,Klein:2013xoa}.    For IceCube, the symmetry is further broken because most of the downgoing neutrinos are produced over Antarctica, where the cold atmosphere leads to slightly increased density, slightly increased neutrino interactions, and therefore a decrease in high-energy neutrinos.

\begin{figure}
\centerline{\includegraphics[width=0.5\textwidth]{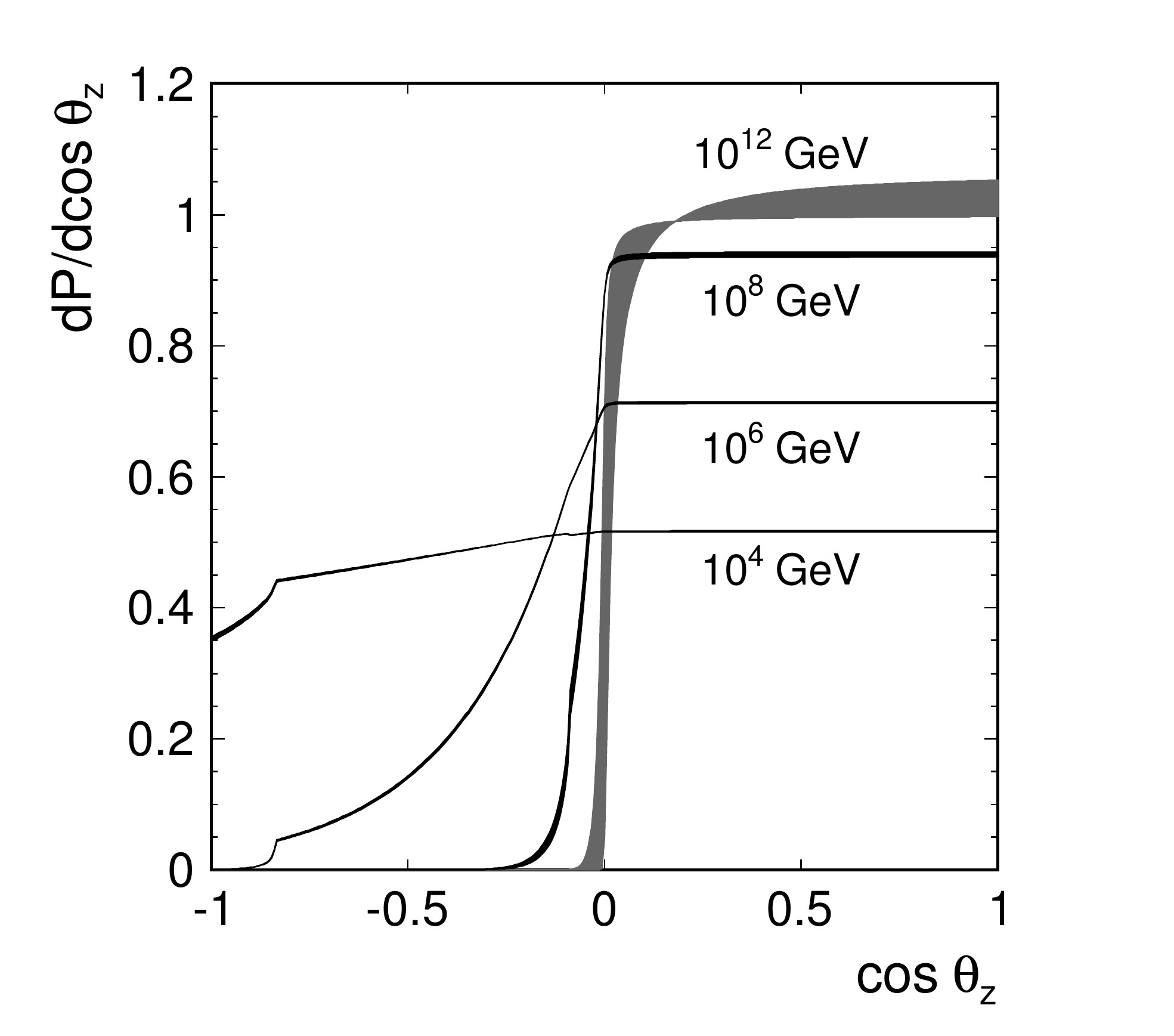}}
\caption{Predicted zenith angle distribution of neutrinos observed in a detector, after passage through the Earth, as a function of energy for the standard-model cross-sections.  With increasing energy, the Earth becomes increasingly opaque, and neutrinos are only observed near the horizon.  Here, $\cos(\theta_z)=1$ corresponds to vertically downward-going, while $\cos(\theta_z)=-1$ corresponds to upward-going.  The kin near $\cos(\theta_z)=-0.8$ corresponds to the Earth's core-mantle boundary; the core is mostly iron, so absorbs more strongly than the rest of the Earth.  From Ref. \cite{Connolly:2011vc}.}
\label{fig:zenith}
\end{figure}

The expected absorption depends on the cross-section.   Figure \ref{fig:zenith} shows the predicted zenith angle distribution for different neutrino energies, assuming the standard model cross section.  The earth is mostly transparent for neutrino energies below 10 TeV, while at energies above about 100 PeV, it is mostly opaque, and neutrinos can only be observed near the horizon.

Recently, it has been seen that the uncertainties on the cross-section are probably larger than on the Earth's density profile.  The circa-1981 Preliminary Earth Reference Model has aged well, showing good agreement with more modern data, with, for long chords through the earth, uncertainties of a few percent \cite{Kennettxx}.   Density studies are still of interest, but they may need to focus on lower energy neutrinos, where the cross-section uncertainties are smaller.  Current tomographic studies suffer from rather large uncertainties \cite{Donini:2018tsg}. One can also use neutrino oscillations to probe the matter density in the Earth; matter induced oscillations depend on the electron density along the path, although there are many complications in going from data to a tomographic model \cite{Winter:2015zwx}.
 
\subsection{$\nu_\mu$ measurements}

A recent IceCube neutrino cross-section measurement \cite{Aartsen:2017kpd}  used one year data, with 79 strings active.  The analysis used a sample of 10,784 upward-going $\nu_\mu$ with measured muon energy $E_\mu> 1$ TeV.    The muon energies were measured using the truncated mean method. The muon tracks were divided into 125 m long segments, and the specific energy loss ($dE/dx$) in each segment was measured. The $dE/dx$ in each segment was determined from the Cherenkov light that was observed  near the track in that segment.  The light emission was assumed to be proportional to the energy loss in the segment.    The 40\% of the segments with the highest measured $dE/dx$ were deleted, and a truncated mean determined from the other segments  \cite{Abbasi:2012wht}.  Although throwing out the segments with the highest energy loss may seem counterintuitive, it leads to smaller uncertainty on the estimated muon energy.  The reason is that the $dE/dx$ distribution has a very large high-side tail, and throwing out these stochastic fluctuations reduces the uncertainty.   The data was binned in terms of muon energy and cos(zenith angle).  Muon energy was used instead of neutrino energy since it was more directly observable.  

 \begin{figure}
\centerline{\includegraphics[width=0.25\textwidth,angle=270]{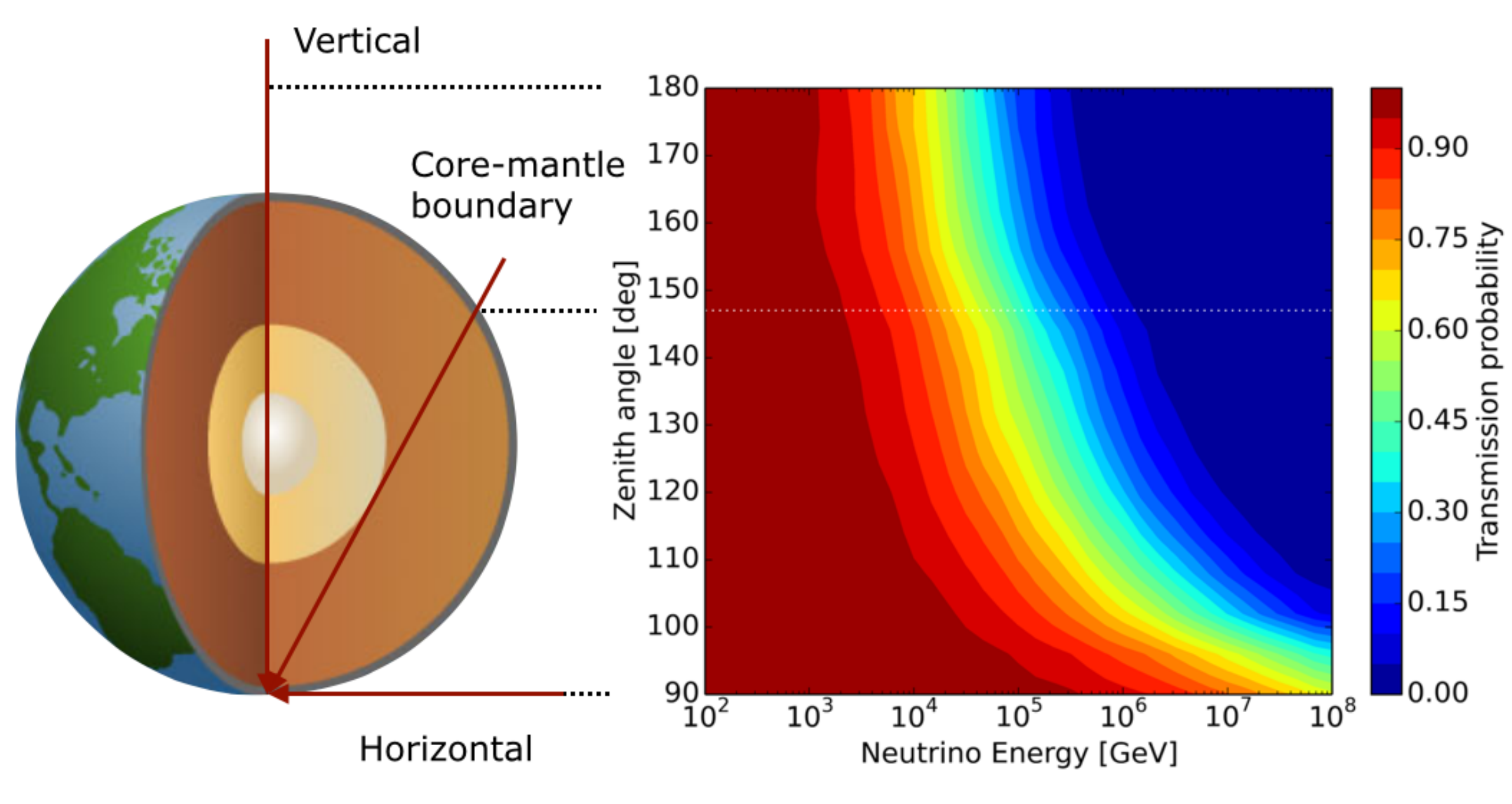}}
\caption{Neutrino absorption changes the zenith-angle dependence of the neutrino energy spectrum.  High-energy neutrinos transiting deep through the Earth are absorbed, while low-energy neutrinos are not.  Neutrinos from just below the horizon provide a nearly absorption-free baseline at all relevant energies.  The right-hand side shows the transmission probability predicted by the Standard Model for neutrinos to transit the Earth as a function of energy and zenith angle.  Neutral current interactions, which occur about 1/3 of the time are included.  When a neutral current interaction occurs, a neutrino is replaced with a neutrino with lower energy.  The horizontal  dotted white line corresponds to a neutrino trajectory that just touches the core-mantle boundary. Adapted from Ref. \cite{Aartsen:2017kpd}.}
\label{fig:principle}
\end{figure}

Figure \ref{fig:principle} shows the idea behind the fit, showing the standard-model transmission probability as a function of neutrino energy and zenith angle.  The binned data was fit by a cocktail of conventional and prompt atmospheric neutrinos, plus astrophysical neutrinos.  The input cocktail included floating parameters to account for uncertainties on the flux magnitudes and spectral indices.   These fluxes were propagated through the Earth, accounting for the neutral-current interaction feed-through using a two-dimensional matrix which depended on the neutrino cross-sections.  The analysis made some simplifying assumptions in order to make the measurement feasible.  The biggest was to assume that the actual cross-sections were an unknown multiple of the CSMS standard model cross-section
\cite{CooperSarkar:2011pa}, $R=\sigma(E_\nu)/\sigma_{\rm SM}(E_\nu)$, with the same multiple holding at all energies.  

In addition to the varying $R$, cross-section, the fit used as input eight nuisance parameters: the atmospheric, prompt and astrophysical flux, a small uncertainty in the index of the cosmic-ray energy spectrum, the $K/\pi$ and $\overline\nu:\nu$ ratio in the conventional cosmic rays, the astrophysical flux power law, and a scaling factor for the uncertainty in the overall sensitivity of the detector optical modules.  Instead of using the fluxes as direct input parameters, it used the $R$ times the flux.  This was done because the $\nu$ flux priors were based on calculations which were tied to previous measurements.  Those analyses implicitly assumed the standard model cross-sections; if the cross-section were doubled, then the flux would be halved.  Fitting for $R$ times the flux removes that connection.   

For each parameter, an initial guess (prior) was assumed, along with an estimated (Gaussian) uncertainty based on our prior knowledge.  The fit returned the most likely result, along with an estimate of the uncertainties.  These parameters are given in Table \ref{tab:sigma}, along with the initial priors and uncertainties.  

\begin{table*}
\caption{Fit parameters with their baseline or units (second column) for the IceCube neutrino cross-section measurement, along with the prior assumption (initial value) and uncertainty input to the fit (third column) and the values returned by the fit (last column).   The neutrino fluxes are for $\nu_\mu$ and $\overline\nu_\mu$ only. For the astrophysical component, the baseline flux is $\Phi_{\rm astro.}\times 10^{-18}(E_\nu/100\ {\rm TeV})^{-\gamma}\ $  s$^{-1}$ cm$^{-2}$ sr$^{-1}$.  The three flux terms are multiplied by $R$ to remove the obvious correlation that the number of observed events increases linearly with the cross section, even in the absence of absorption.}
 {\begin{tabular}{llcrr}
Result& Baseline/units   & Nuisance Parameter & Nuisance Parameter \\
        &            &  Input $\&$ uncertainty $\sigma$       & Fit result  \\
         \hline
$\Phi_{\rm Conv.}\times\sigma$ & Ref.  \cite{Honda:2006qj} $\times  R$ ($R=\sigma_{\rm meas.}/\sigma_{\rm SM}$) &  $1.0 \pm 0.25$      & $0.92 \pm 0.03$ \\
 $\Phi_{\rm Conv.}$  spectral index    &   Ref. \cite{Honda:2006qj} with knee   &   $0.00 \pm 0.05$          &  $+ 0.007\pm 0.001$ \\
 K/$\pi$ ratio				     &  Ref. \cite{Honda:2006qj} baseline               & $1.0\pm 0.1$           & $1.05\pm0.09$ \\
 $\nu/\overline\nu$ ratio                 &    Ref. \cite{Honda:2006qj} baseline               & $1.0\pm 0.1$       & $1.01 \pm 0.005$   \\
$ \Phi_{\rm prompt}\times\sigma$   & Ref.  \cite{Enberg:2008te}$\times R$           & $0.0_{-0.0}^{+1.0}$ & $0.5^{+0.40}_{-0.34}$ \\
 $\Phi_{\rm astro.}\times\sigma$ &     Ref. \cite{Aartsen:2015knd}$\times R$         & $2.23\pm 0.4$  & $2.62_{-0.07}^{+0.05}$ \\
 Astrophysical index ($\gamma$)    &                                                                         & $2.50\pm 0.09$   &  $2.42\pm 0.02$ \\
 DOM Efficiency		               & IceCube Baseline                                                  & $1.0\pm 0.1$     & $1.05\pm 0.01$ \\
\hline
 \end{tabular}}
 \label{tab:sigma}
 \end{table*}

The fit finds $R=1.30^{+0.30}_{-0.26}$.  The other (nuisance) parameters are compatible with their input priors, within the uncertainties.  Most of this was statistical error, but some was due to uncertainties in nuisance parameters.  These were separated by redoing the fit, with the nuisance parameters fixed to their preferred values; this gave the statistical errors, and the systematic errors associated with the fit were determined by quadratic subtraction.  

Some additional systematic errors  could not be easily included in the fit, so they were evaluated separately.   The largest uncertainty concerned the optical properties (scattering and absorption length) of the ice.  This was evaluated by running the analysis with two different ice models, and comparing the results.   Other systematic errors came from uncertainties in the density distribution of the Earth, variations in atmospheric pressure at the neutrino production sites (this can impact the angular distribution of conventional atmospheric neutrinos), and uncertainties in the optical angular acceptance of the DOMs.  IceCube added one additional systematic uncertainty, related to the spectral index of the astrophysical flux.   Previous analyses had measured different astrophysical spectral indices, with cascade or contained event studies finding $\gamma\approx 2.5$, and studies based on through-going muons finding lower number, $\gamma\approx 2.1$.  The collaboration decided to use the former number, because it was obtained by a method very different from the cross-section analysis.  An additional systematic error was included to allow for the larger total uncertainty. 

\begin{figure}[t]
 \centerline{\includegraphics[width=0.48\textwidth,angle=0]{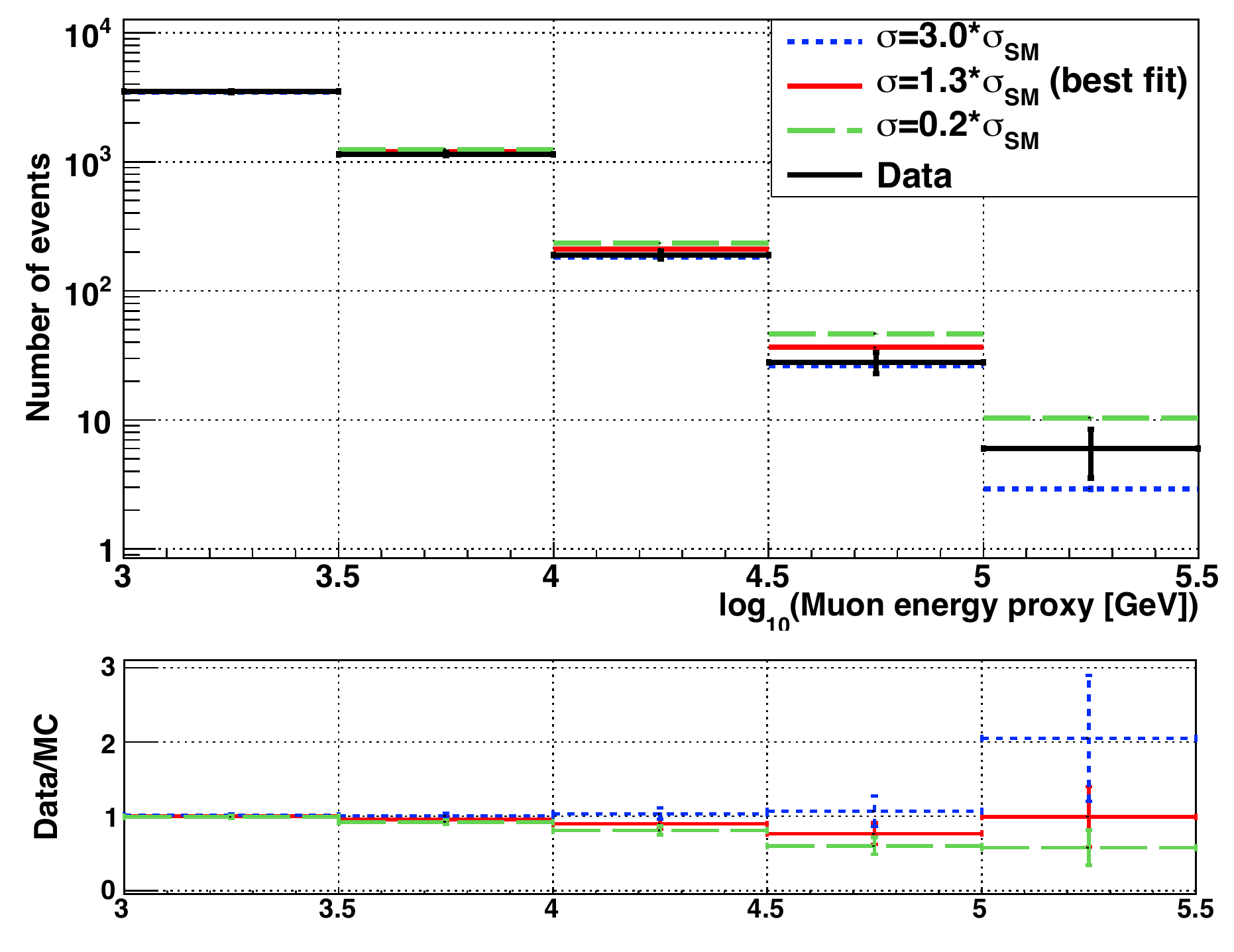}}
\caption{The IceCube $\nu_\mu$ data, compared with three Monte Carlo model predictions.  The black lines with error bars show the muon energy distribution for muons with zenith angles between 110 and 180 degrees, while the dashed green, solid red and dotted blue lines show predictions for cross-sections with $R=0.2$, $R=1.3$ (the best fit) and $R=3.0$ respectively.   The bottom panel shows the ratio of the data to the three predictions.  The three predictions agree at low energies, where neutrino absorption is negligible, but diverge at higher energies, with the $R=1.3$ points best matching the data.  The data covers the angular range where neutrino absorption is significant.   
From Ref. \cite{Aartsen:2017kpd}.}
\label{fig:fits}
\end{figure}

With these uncertainties, IceCube measured a cross-section of $R=1.30^{+0.21}_{-0.19} ({\rm stat.}) ^{+0.39}_{-0.43} ({\rm syst.})$ times the standard model.   Figure \ref{fig:fits} compares the data with simulations where the cross-section was fixed to three different values, from 0.2 to 3.0 times the standard model.   The two extrema are clearly excluded, while the $R=1.3$ gives a match between the model and the data.  

Determining the energy range where such an analysis is valid is not straightforward.  The lower limit comes at the point where absorption is negligible, while the upper limit occurs when the analysis runs out of statistical power.  Different ways to quantify these points lead to surprisingly different answers.  In the end, to find the lower end of the sensitivity range, IceCube made a series of fits where they turned off absorption up to an energy threshold, and studied how the likelihood of the fit increased.  The minimum sensitive energy was taken to correspond to a likelihood increase of $-2\Delta LLH$=1 (a $1\sigma$ change if Gaussian statistics is assumed).  A similar procedure was followed to find the maximum sensitive energy, turning off absorption above a threshold energy, and again finding the point at which $-2\Delta LLH$ increased by 1.   This gave a wide energy range of 6.3 TeV to 980 TeV.  The wide range is partly due to the nature of the problem - the increasing cross-section, combined with the decreasing number of events with increasing energy naturally leads to a wide range.  But, some of it is due to the method; the $-2\Delta LLH=1$ criteria gives the entire range over which the analysis has some sensitivity.  One drawback of the method is that  the result depends on the amount of data; using the same technique with a longer detector exposure would lead to a still wider range.

\begin{figure}
\centerline{\includegraphics[width=0.5\textwidth]{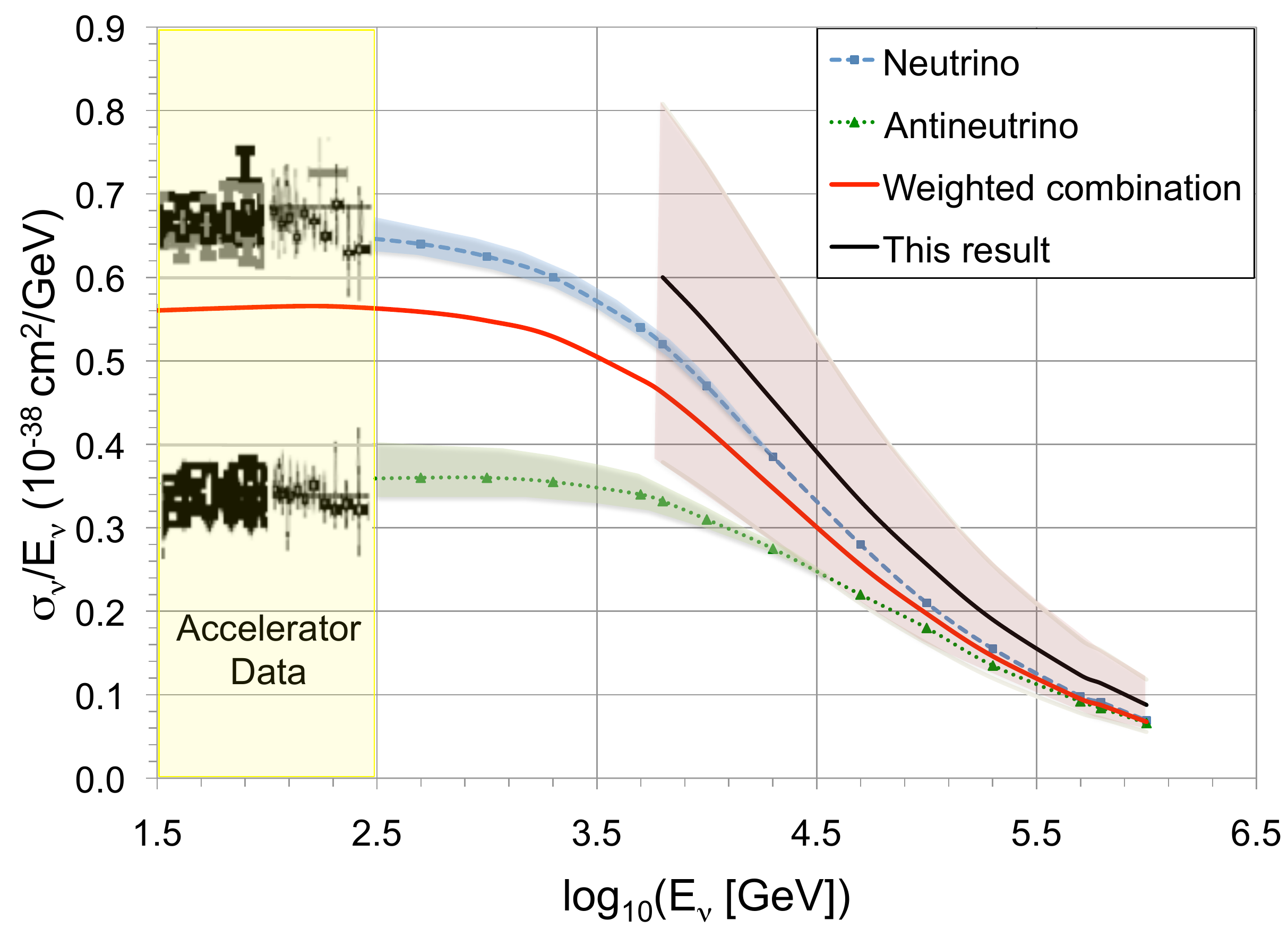}}
\caption{The IceCube cross-section measurement, compared with previous measurements made at accelerator laboratories.  From Ref.
\cite{Aartsen:2017kpd}.}
\label{fig:results}
\end{figure}

Figure \ref{fig:results} shows the IceCube results, in terms of cross-section divided by energy.  This metric is chosen because otherwise a logarithmic y-axis would be needed.  The $\nu N$ dotted blue curve (and upper data points) is proportional to energy up to about 1 TeV, then rises more slowly, roughly as $E_\nu^{0.3}$.  The reduction in slope is due to the finite masses of the $W^\pm$ and $Z^0$.  

 \subsection{Cascade/contained event measurements}
 
The cross-section can also be measured using cascades and/or starting events; the latter is a broader category including $\nu_\mu$ interactions that occur within the detector.   Bustamante and Connolly have analyzed public IceCube data on contained events \cite{Bustamante:2017xuy} and  a preliminary IceCube analysis compares the rates of upward and downward-going events \cite{Aartsen:2017mau}.

\begin{figure}[t]
 \centerline{\includegraphics[width=0.45\textwidth,angle=0]{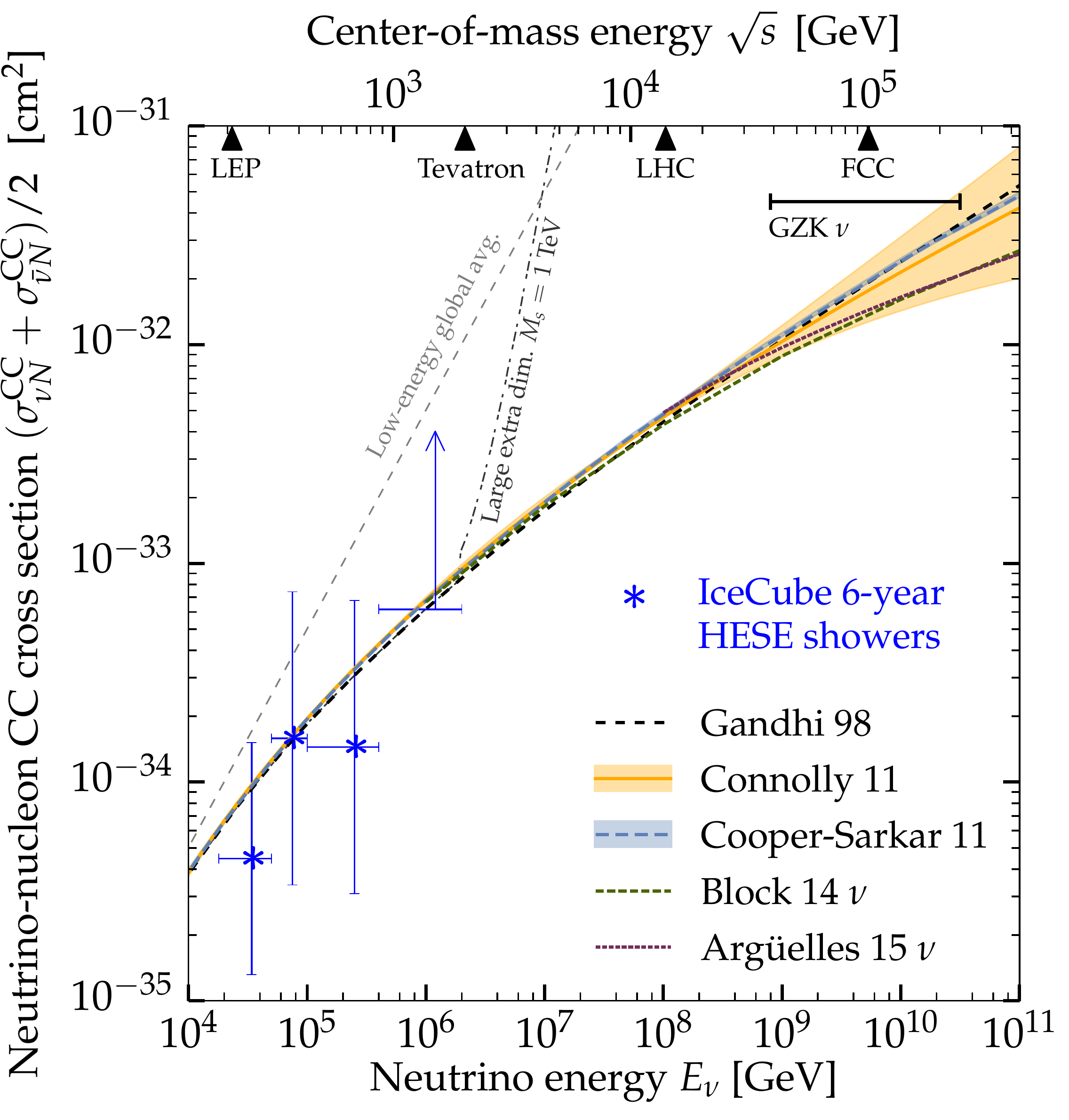}}
\caption{Different predictions for neutrino CC cross-sections (lines), compared to measurements based on contained IceCube events.  At low energies, the cross-section increases linearly with energy.  The plot also shows the predictions for one model involving extra dimensions \cite{AlvarezMuniz:2001mk}.   Also shown at the top are points showing the center-of mass energy reach from three particle colliders.  From  Ref. \cite{Bustamante:2017xuy}.}
\label{fig:bcsigma}
\end{figure}

Cascades/contained events have both advantages and disadvantages compared to using through-going muons from $\nu_\mu$.  One big disadvantage is that the data samples are smaller.  The authors in Reference~\cite{Bustamante:2017xuy} used 58 contained events from 6 years of IceCube data, vs. the 10,784 from one year in the original IceCube $\nu_\mu$ search.  A second disadvantage is that the zenith angle uncertainty for cascade events is large, often 10-15 degrees, and it may be subject to systematic pulls due to uncertainties in optical scattering in the ice.  Third, cascades include both charged-current $\nu_e$ and $\nu_\tau$ events, along with neutral-current interactions.  This is a particularly large issue at lower energies, where atmospheric $\nu_\mu$ dominate, leading to a particularly large neutral current contribution which mixes sensitivity to the different neutrino flavors.   On the other hand, the neutrino energy is well known, so that it is easy to divide the analysis up in energy bins.     Figure \ref{fig:bcsigma} shows the results from reference~\cite{Bustamante:2017xuy}, presented in 4 energy bins, from 18 TeV, going up to 2 PeV.  The analysis has many similarities to the IceCube analysis, with a cocktail of atmospheric and astrophysical neutrinos. It assumes that the inelasticity distribution follows the standard model, since a model for this is needed to convert the energy deposited in the detector into neutrino energy.

The upward:downward IceCube analysis uses a larger sample, based on 5 years of cascade interactions \cite{Aartsen:2017mau}.  It has similarities with Bustamante and Connolly, but, since it uses just two bins in zenith angle (upgoing and downgoing), it has reduced sensitivity.  Both analyses benefit from the good cascade energy resolution, which allows them to measure the cross-section in multiple energy bins. 

\subsection{Future cross-section measurements}

Future analyses should make a wider range of measurements.   With more data, future $\nu_\mu$ measurements should be able to fit for multiple $R$s, each covering a different energy range, even with the poor energy resolution.  Starting tracks are also sensitive to the cross-section; they offer both good angular resolution (hence good path-length determination) and reasonably good energy resolution, with reasonable event sizes.

Future analyses can also consider a broader range of hypotheses.  The assumption that the CC and NC cross-sections change in tandem is appropriate if one assumes that the interactions are governed by the standard model, where most of the uncertainty comes from the parton distributions.    However, many models involving new physics do not have a neutral current counterpart - if the neutrinos interact, they  disappear.  So, to test for new physics, it may be best to vary only the CC cross-section.  Ideally, analysts will do model tests, making fits to test if particular models are ruled out.   These tests can account for specific predictions about how the cross-section varies with energy.   These tests can also account for any neutrino emission (at a lower energy than the incident neutrino) from interactions, as predicted by that specific model. 

\section{Interaction characteristics and inelasticity}

Any new interaction that adds to the cross-section is likely to produce final states that differ from those that are produced in standard-model interactions.  So, another way to search for new types of interactions is to look at the final states of neutrino interactions that occur within a detector.   There are also a few possible interesting signatures that come from interactions that occur far outside of the detector.

The neutrino inelasticity is the fraction of the neutrino energy that is transferred to a nuclear target.  To measure the  inelasticity, it is necessary to measure both the energy of the hadronic shower from the nuclear target, and the energy of the outgoing lepton.  This is only possible for $\nu_\mu$ that interact within the detector. One key difficulty in this kind of analysis is separating the light that comes from the shower and the light that comes from the muon, plus  making sure that the event contains an outgoing muon, and is not just a cascade.  The need to see a muon track and measure its energy limits the acceptance at very high inelasticity, particularly near threshold.

\begin{figure*}
\centerline{\includegraphics[width=0.9\textwidth]{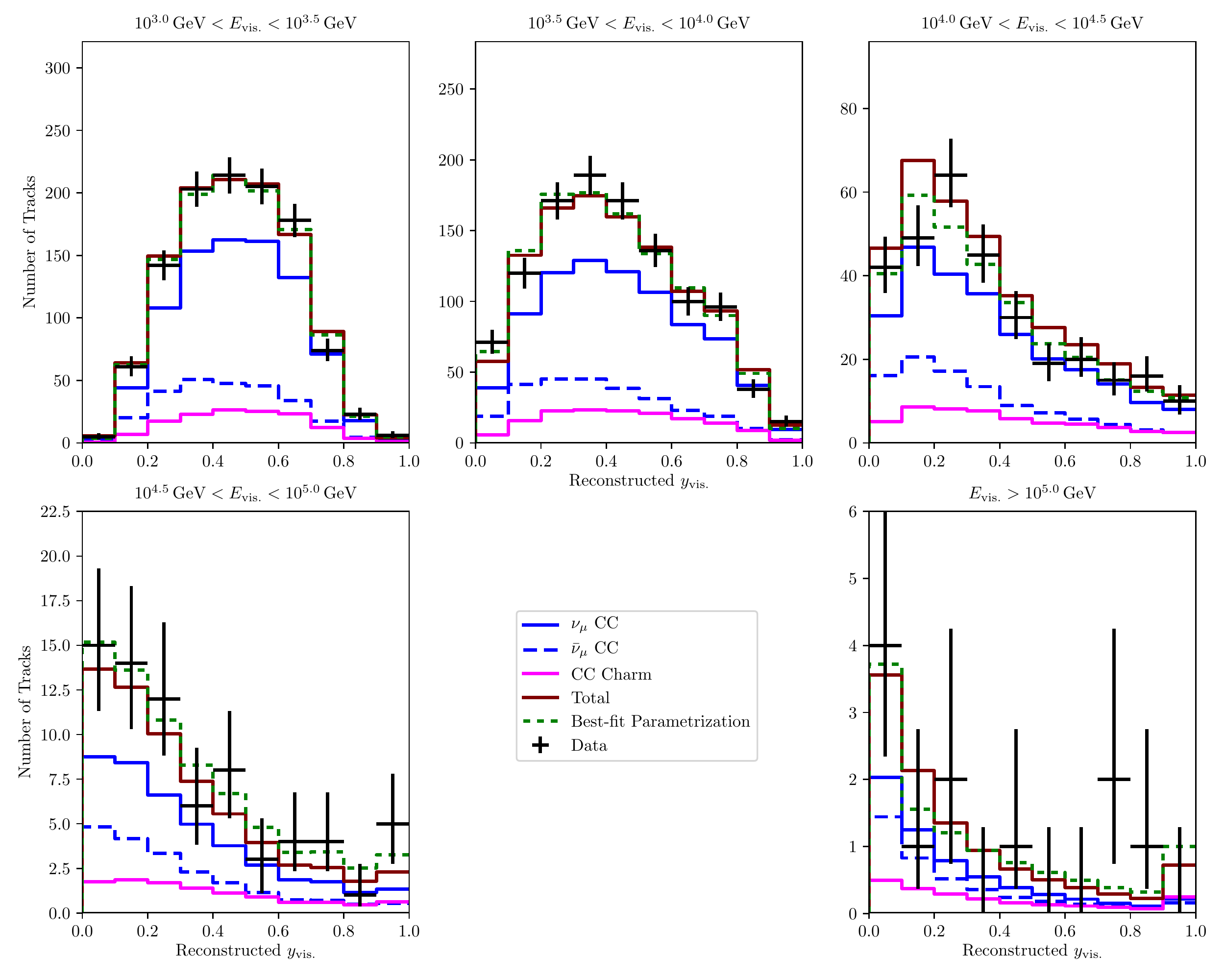}}
\caption{Inelasticity distributions measured in IceCube starting track events, in five energy bins. The $y$ axes show inelasticity, while the 5 plots are for different energy ranges.
From Ref. \cite{Aartsen:2018vez}.}
\label{fig:inelasticity}
\end{figure*}

IceCube has made such a measurement using a sample of 2,650 contained starting track events from 5 years of data \cite{Aartsen:2018vez}.  The central 68\% of the track events had estimated neutrino energies between 11 TeV and 410 TeV.  A parallel sample of 965 cascades was also used to help constrain the fits; it had 68\% central energy range of 8.6 to 207 TeV. A machine learning algorithm was used 
to separate the track and pure cascade events, and another to determine the track and cascade energies.  Most of the tracks were energetic enough to travel far enough to leave the detector, so most of the muon energy estimate came from the muon specific energy loss ($dE/dx$).   The energy resolution was quantified in terms of the logarithm of the energy:  log$_{10}(E_\mu/1\ {\rm GeV}) = 0.36$ and log$_{10}(E_{\rm casc.}/1 {\rm GeV}) =0.19$ for the associated cascades.  Because of an anti-correlation, the neutrino energy resolution was better than either the track or the cascade resolution, log$_{10}(E_{\rm vis.}/{\rm 1 GeV})=0.18$.   The visible inelasticity, $y_{\rm vis}$ was defined as
\begin{equation}
y_{\rm vis} = \frac{E_{\rm casc}}{E_{\rm track} + E_{\rm casc.}}.
\end{equation}
The visible inelasticity is somewhat different from the actual inelasticity because of measurement uncertainties, plus some biases.  The main biases are the loss of cascade energy due when it emits neutrinos, and the difference in light output for hadronic showers compared to electromagnetic.   IceCube performed fits to the visible inelasticity, comparing data distributions with theoretical predictions that have been filtered through the IceCube simulations and reconstruction.  The $y_{\rm vis}$ distributions found with this method are shown in Fig. \ref{fig:inelasticity}.   
$y_{\rm vis.}$ is plotted in four half-decade energy bins, from 1 TeV to 100 TeV, with  a 5th bin for energies above 100 TeV.   The data is compared with the expectations from a fit which includes $\nu_\mu$ and $\overline\nu_\mu$ interactions, along with a separately treated charm component, all following the CSMS inelasticity calculation.  The sum is in good agreement with the data. 

Because of the limited acceptance (especially at very large or very small $y$), it is not straightforward to convert the $y_{\rm vis.}$ distribution into a $y$ distribution.   Instead, IceCube parameterized the distribution, starting from the lowest order cross-sections
\begin{equation}
\frac{d\sigma}{dxdy}=
 \frac{2 G_{F}^{2}M E_{\nu} }{\pi}\left(\frac{M_{W}^{2}}{Q^{2}+M_{W}^{2}}\right)^{2} \\ \big[x q(x,Q^2) + (1-y)^2 x \bar{q}(x,Q^2)\big],
  \label{eq:nu_cc_xs_xy_param}
\end{equation}
where $q$ and $\bar{q}$ are the summed  quark parton distribution functions (PDFs).  At low values of Bjorken-$x$, where they should have a power-law behavior, $xq(x,Q^2)\sim A(Q^2)x^{-\lambda}$ with $\lambda \sim 0.4$.  When transforming variables from $(x,y)$ to $(Q^2,y)$, the $Q^2$-dependence of Eq.~\ref{eq:nu_cc_xs_xy_param} can be separated from the $y$-dependence and integrated out to give a two-parameter function, 
\begin{equation}
  \frac{d\sigma}{dy} \propto \left(1 + \epsilon (1-y)^2\right)y^{\lambda - 1}.
  \label{eq:nu_cc_xs_y_param}
\end{equation}
 Here $\epsilon$ sets the relative importance of the term proportional to $(1-y)^2$ in Eq.~\ref{eq:nu_cc_xs_xy_param}.  The incident `beam' consists of a mixture of neutrinos and antineutrinos, which have different predicted $\epsilon$.  The measurement gives the weighted average.  The  inelasticity distribution is \begin{equation}
  \frac{dp}{dy} = N \left(1 + \epsilon (1-y)^2\right)y^{\lambda - 1},
  \label{eq:nu_cc_dpdy_param}
\end{equation}
where $N$ is the normalization
\begin{equation}
  N = \frac{\lambda(\lambda+1)(\lambda+2)}{2\epsilon + (\lambda+1)(\lambda+2)}.
\end{equation}
Unfortunately, in a fit to Eq. \ref{eq:nu_cc_dpdy_param}, $\epsilon$ and $\lambda$ are highly correlated.  To avoid this correlation, the collaboration instead fit for $\langle y\rangle$ and $\lambda$, which exhibit much less correlation, where $\langle y\rangle$ is defined as:
\begin{equation}
  \langle y \rangle = \int_0^1 y \frac{dp}{dy}dy = \frac{\lambda (2\epsilon + (\lambda + 2)(\lambda + 3))}{(\lambda + 3)(2\epsilon + (\lambda + 1)(\lambda + 2))}.
  \label{eq:meany}
\end{equation}
Then
\begin{equation}
  \epsilon = -\frac{(\lambda+2)(\lambda+3)}{2}\frac{\langle y \rangle(\lambda + 1) - \lambda}{\langle y \rangle(\lambda + 3) - \lambda}
\end{equation}
and $dp/dy$ can easily be found as a function of $\langle y\rangle$ and $\lambda$ only.

\begin{figure}
  \centering
  \includegraphics[width=0.9\linewidth]{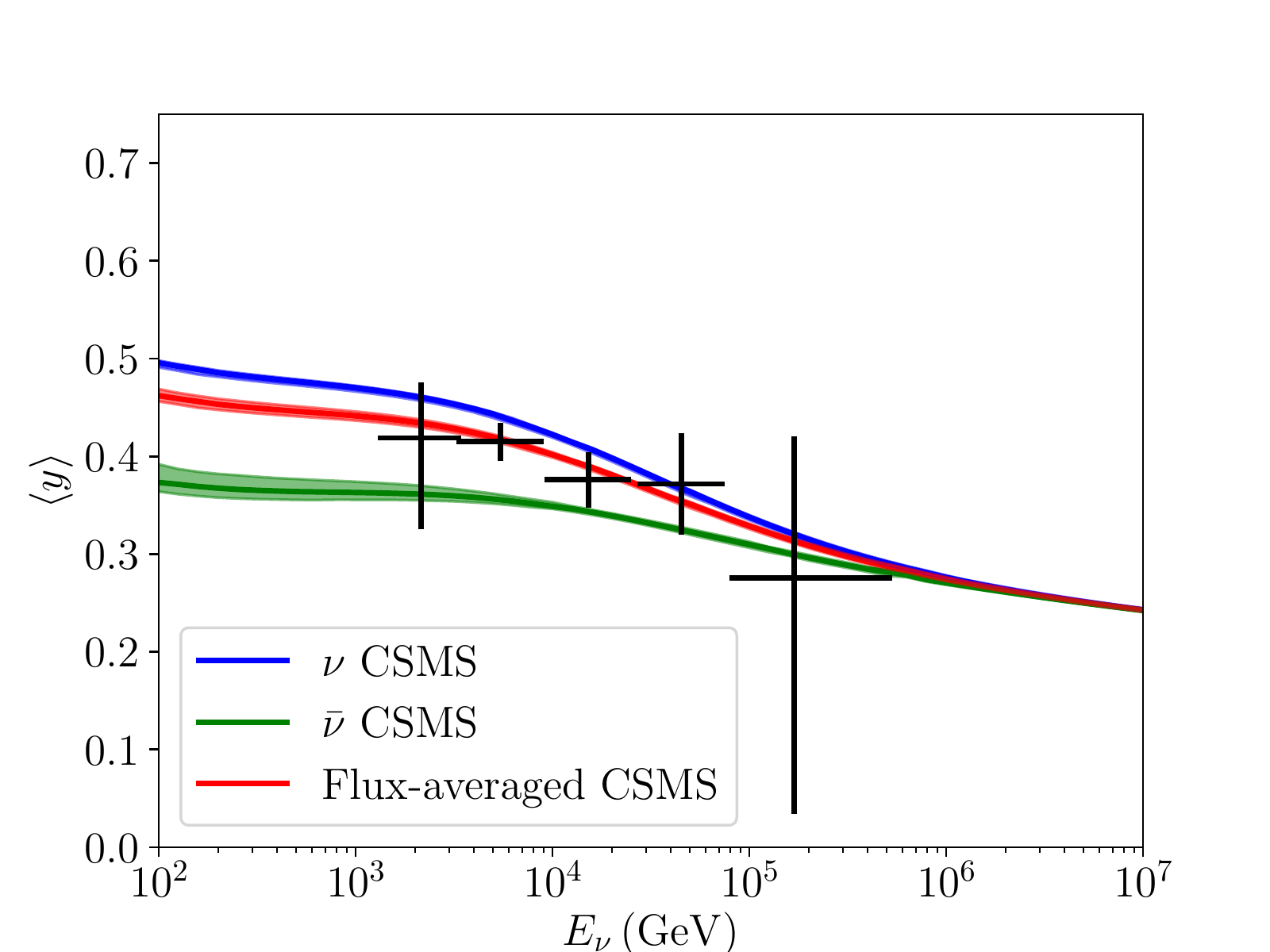}
  \caption{The mean inelasticity obtained from the fit to Eq.~\ref{eq:meany} in five bins of reconstructed energy.  Vertical error bars indicate the $68\%$ confidence interval for the mean inelasticity, and horizontal error bars indicate the expected central $68\%$ of neutrino energies in each bin.  The blue and green lines show the predicted mean inelasticity from CSMS for neutrinos and antineutrinos respectively, while the red line shows the expected average, based on the estimate for atmospheric neutrinos and assuming an even $\nu/\overline\nu$ split for astrophysical neutrinos. From Ref. \cite{Aartsen:2018vez}.}
\label{fig:split_fit_inel}
\end{figure}

Figure \ref{fig:split_fit_inel} shows $\langle y\rangle$ as a function of energy.  $\langle y\rangle$ drops with increasing energy, in agreement with the CSMS calculation.  The green and blue lines show the expected distributions for $\nu$ and $\overline\nu$ respectively, while the red line shows the prediction for the best-fit mixture of atmospheric and astrophysical neutrinos described in  Sec.~\ref{sec:sources}. The choice of flux model has little effect on the inelasticity distribution due to the relatively narrow width of the energy bins.  The data are in good agreement with the predictions.

The charm component is treated separately in Fig.  \ref{fig:inelasticity}  because neutrino interactions that produce charm have a different inelasticity distribution from interactions involving only light quarks.  There are two reasons for this.  First, charm quarks are produced when a neutrino interacts with a strange sea quark in the target.  There are no strange valence quarks, so the inelasticity distribution is flatter, and the abundance of high $y$ events is sensitive to charmed quark production; this can be a way to measure the $s$ quark distribution in nucleons.  The charm-production cross-section is predicted to rise slowly, from 10\% of the cross-section at 100 GeV, to 20\% at 100 TeV. 

Second, about 10\% of charmed hadrons decay semi-leptonically, producing a muon or an electron.  These muons will be indistinguishable from the main muon, and add to the total observed muon energy.   IceCube made a fit where the charmed cross-section was allowed to be a varying fraction of the standard model prediction, and found that the cross-section was $0.93^{+0.73}_{-0.59}$ times the standard model cross-section.  A likelihood analysis found that charm production was observed with more than 90\% confidence level.   The fit was sensitive in the neutrino energy range 1.5 to 340 TeV.

IceCube also used the inelasticity distribution to measure the $\nu/\overline\nu$ ratio.  This depends on the different inelasticity distributions for $\nu$ and $\overline\nu$.  As can be seen from Fig. \ref{fig:split_fit_inel},  $\nu$ and $\overline\nu$ have different $\langle y\rangle$  but only for neutrino energies below about 10 TeV.  The measurement is only sensitive to atmospheric neutrinos which dominate at lower energies.   The collaboration added a scaling parameter to the atmospheric $\nu/\overline\nu$ ratio, and found that the fitted parameter was compatible with predictions \cite{Honda:2006qj}.   Looking ahead, it might be possible to apply this method to astrophysical neutrinos if one could find a way to veto the atmospheric flux.  One approach might be to use only downward-going neutrinos, with a large, low-threshold surface detector array to veto events accompanied by a cosmic-ray air shower.  

IceCube also used the data to fit for the astrophysical neutrino flavor ratio.  By using the inelasticity, the collaboration was able to get significantly tighter bounds than were possible without it.  With additional data, it will be possible to fit the inelasticity distribution directly, to look for a $\nu_\tau$ component. 

\subsection{Other interaction signatures}

Other beyond-standard-model physics could produce other types of interactions, which might produce other signatures in neutrino telescopes.   Here, we  discuss one new signature which could signal supersymmetry or Kaluza-Klein models with extra dimensions.  

In models of supersymmetry where the lightest particle is a gravitino, the next-to-lightest supersymmetric particle (NLSP) is a long-lived, charged slepton \cite{Albuquerque:2006am}.   Supersymmetric particles may be produced in pairs by neutrino interactions in the Earth.  When heavier supersymmetric particles are produced, they will cascade down in a series of decays, eventually becoming a slepton which may be fairly long-lived.  The pair of sleptons will propagate through the Earth.  If they are heavy enough (masses of a few hundred GeV), they will lose energy slowly (as minimum-ionizing particles), and so can propagate for long distances.   The sleptons will both have high momentum, so they will follow very similar trajectories through the Earth, diverging slowly due to their relative transverse momentum and due to multiple scattering in the Earth \cite{Albuquerque:2009vk}; the Earth's magnetic field does not matter much.   A pair of nearly parallel upward-going, minimum-ionizing tracks  passing through a neutrino telescope would be a very distinctive signature.  IceCube is sensitive to pairs with separation from about 135 m to 1 km \cite{Helbing:2011wf,Abbasi:2012kza}.  A preliminary search found no evidence of a signal \cite{Kopper:2015rrp}. The only background, due to a pair of atmospheric neutrinos which both interact simultaneously in the ice or rock below the detector, is small, but could be visible in detectors several times larger \cite{vanderDrift:2013zga}.

A similar scenario applies in some models with universal extra dimensions (Kaluza-Klein models).  In some of these models, the lightest KK particle (LKP) is neutral and interacts extremely weakly, while the next-to-lightest KK particle (NLKP) is charged.   If the coupling between the NLKP and LKP is small enough, the NLKP becomes long-lived, leading to a similar signature.     Unfortunately, searches to date have not found a signal \cite{Aartsen:2015bwa}.  

\section{Future prospects}

There are opportunities for significantly improving all of these measurements in the coming years.    More data will become available from optical Cherenkov detectors.  The current $\nu_\mu$ cross-section analysis, in particular, only  used one year of data; substantial improvements are available by using multiple years of data.  Slightly further ahead, experiments like KM3NeT Phase 2.0 \cite{Adrian-Martinez:2016fdl} and IceCube Gen2 \cite{Aartsen:2014njl} should instrument of order 5 km$^3$ and 10 km$^3$ respectively.    These experiments might push current measurements up to 10 PeV, but the neutrino flux drops rapidly with increasing neutrino energy, and it is hard to imagine an optical Cherenkov detector much bigger than 10 km$^3$.  This limits the usable neutrino flux to around 10 PeV.  At higher energies, there is a new source of neutrinos, the `GZK,' mechanism, due to interactions of ultra-high energy (above about $4\times10^{19}$ eV) cosmic-ray protons with the cosmic microwave background radiation (CMBR) \cite{Abbasi:2007sv}.  The protons are photoexcited to the $\Delta^+$ resonance, which decays, eventually producing a lower energy proton, plus neutrinos and/or photons.   These GZK neutrinos have energies in the $10^{18}$ to $10^{20}$ eV range; the flux depends on the cosmic-ray composition, but they should be visible in a large (order 100 km$^3$) detector.

Fortunately, a new technique may allow us to instrument the required volumes at energies above 10 PeV.  That technique is to observe the radio-frequency coherent Cherenkov emission from the showers that are produced in neutrino interactions \cite{Klein:2010wf,Connolly:2016pqr}.     In ice, showers have a transverse size (Moliere radius) of about 11 cm.  When viewed at wavelengths larger than this scale (more precisely, the radius divided by the cosine of the Cherenkov angle), then we do not see radiation from individual particles.  Instead, the amplitudes for Cherenkov emission add in phase, and the radiation scales as the square of the net charge, {\rm i. e.} as the square of the neutrino energy.  As Gurgan Askaryan pointed out \cite{Askaryan1,Askaryan2}, this net charge is significant, for two reasons.  First, MeV photons in the shower may scatter atomic electrons into the beam.  Second, positrons in the late stages of the shower may annihilate on atomic electrons.  Because of these two effects, the net charge in the shower is about 10\% of the total charge. This leads to a large radio pulse at frequencies up to about 1 GHz. 

Radio detectors observe coherent Cherenkov emission, so are sensitive to concentrated showers, rather than muons.    As Fig. \ref{fig:zenith} shows, at these energies, the Earth is mostly opaque to neutrinos, so the cross-section measurement is done with neutrinos that are coming from near the horizon.  Because of the narrow range of zenith angles, good zenith angle resolution is important.    This angular resolution is achievable, in principle, by combining information on the frequency spectrum of the pulse and its polarization.  The frequency spectrum gives the angular distance between the receiver-interaction vector and the Cherenkov cone, while the polarization gives the perpendicular angle, the angular distance between the receiver and the neutrino direction.  Simulations show that this can give angular resolutions of a few degrees.  

Most current experiments have had thresholds near or above $10^{20}$ eV, and have not yet observed neutrinos \cite{Patrignani:2016xqp}.  The high thresholds are because the antennas are well separated from the active volume.   The only positive claims of neutrino observation are by the ANITA balloon-borne experiment, which has reported two anomalous events which may be interpreted as being $\nu_\tau$ \cite{Gorham:2018ydl}.  The events are anomalous because their energy, about $10^{17}$ eV is high enough so that they should have been absorbed while traversing the Earth.  Many beyond-standard-model explanations have been put forth to explain the events, but more prosaic explanations should also be further considered.   

Future efforts are focused on deploying antennas in or near the Earth's surface, to better cover this lower energy range.   Two collaborations, ARA at the South Pole \cite{Allison:2015eky}, and ARIANNA on the Ross Ice Shelf \cite{Barwick:2014pca} have deployed modern prototype detectors with antennas directly in the active volume, with thresholds below $10^{17}$ eV.  ARA has also demonstrated a phased-array (interferometric) trigger, which can reach down to $10^{16}$ eV \cite{Allison:2018ynt}, and so overlap, or nearly overlap with optical Cherenkov measurements.   Looking ahead, portions of the two groups have proposed two new arrays, Radio Neutrino Observatory (RNO) \cite{Buson:2019dbj} and Askaryan Radio In-ice Array (ARIA) \cite{Anker:2019mnx} for deployment at the South Pole.  Both would study GZK neutrinos, while RNO would use a phase array trigger to lower its threshold, and study astrophysical neutrinos down to about $10^{16}$ eV.  They hope to deploy in the early 2020's. 

Both experiments might expect to observe a few handfuls of events.  Depending on the zenith angle resolution, this may be marginal for cross-section measurements, which seem to require of order 100 events for a useful (factor of 2) measurement.  However, a future larger radio array could make such a measurement.  There are some possibilities on the horizon.   IceCube Gen2 may include a larger radio array \cite{Aartsen:2014njl}, notionally 5 times larger than RNO.  Or, the GRAND collaboration is proposing to deploy an enormous array - up to  in 300,000 antennas - in Eastern China \cite{Alvarez-Muniz:2018bhp}.   GRAND would detect cosmic-rays, and also the showers from $\tau$ lepton decays; the $\tau$ are produced when upward-going $\nu_\tau$ interact in the rock below the detector, and can exit the Earth before decaying.   GRAND faces many challenges in rejecting anthropogenic and other backgrounds, but either detector could make a cross-section measurement. 

This measurement would be sensitive to the beyond-standard-model processes discussed in Section \ref{sec:theory} at much higher mass scales, above those that are accessible at CERNs Large Hadron Collider.  It would also provide a measurement of standard-model parton distribution at Bjorken$-x$ values down to $10^{-6}$ at large $Q^2$, thereby testing models of parton saturation.

IceCube Gen2 and KM3NeT 2.0 should be able to extend the inelasticity measurements upward in energy, to at least a PeV, enough to probe for new physics at a moderate mass scale.   Because the radio-detection technique is only sensitive to showers, it will have difficulty with inelasticity measurements.     But, at the highest energies, the Landau-Pomeranchuk-Migdal (LPM) effect \cite{Klein:1998du} may come to the rescue.  The LPM effect reduces the cross-sections for bremsstrahlung and pair production, so elongates electromagnetic showers, producing a series of subshowers downstream of the neutrino interaction site \cite{CastroPena:2000fx,Gerhardt:2010bj}.  Hadronic showers are much less subject to this effect, so will remain clustered at the interaction point.  A sufficiently granular radio array might be able to separate the main shower from the subshowers, and thereby determine the fraction of the neutrino energy that was transferred to the target.   Phase array techniques may be particularly helpful in that regard. 

\section{Conclusions}

Neutrino telescopes offer an opportunity to study neutrino physics at energies up to a few PeV, many orders of magnitude higher than are available at accelerators.  Studies have been made of both the neutrino cross-sections, and also the final states produced, specifically the neutrino inelasticity distribution.  
These measurements are sensitive to many aspects of standard-model neutrino physics, and are also sensitive to some beyond-standard-model physics.   
Future measurements with radio-Cherenov detectors should reach energies above $10^{17}$ eV, enough to probe new physics at energies beyond those accessible at the LHC.   

Although the energy reach is impressive, the limited neutrino flux and large detector granularity limit the precision of current measurements.  However, they are our only probe of neutrino energies above 1 TeV, and provide a useful new window for neutrino physics reaching up to and beyond LHC energies.

\section{Acknowledgements}

I thank Drs. Sandra Miarecki and Gary Binder for their respective dissertation work on the IceCube $\nu_\mu$ cross-section and inelasticity analyses. This work was supported in part by the National Science Foundation under grant number PHY-1307472 and the U.S. Department of Energy under contract number DE-AC- 76SF00098.

\bibliography{book}


\end{document}